\documentstyle[prb,aps,psfig]{revtex}
\begin{document}

\draft

\title{Ground-state properties of rutile:
electron-correlation effects}

\author{Krzysztof Ro\'sciszewski, Klaus Doll, Beate Paulus, Peter Fulde}

\address{Max Planck Institut f\"{u}r Physik komplexer Systeme,
N\"{o}thnitzerstr. 38, D-01187 Dresden, Germany}

\author{Hermann Stoll}

\address{Institut f\"{u}r Theoretische Chemie, Universit\"{a}t
Stuttgart, D-70550 Stuttgart, Germany}

\maketitle

\begin{abstract}
Electron-correlation effects on cohesive energy, lattice constant and bulk 
compressibility of rutile are calculated using an {\em ab-initio} scheme.
A competition between the two groups of partially covalent Ti-O bonds
is the reason that the correlation energy does not change linearly with
deviations from the equilibrium geometry,
but is dominated  by quadratic terms instead.
As a consequence, the Hartree-Fock lattice constants are
close to the experimental ones, while
the compressibility 
is strongly renormalized by electronic  correlations.
\end{abstract}

\narrowtext

\section{Introduction}
Although transition-metal oxides are one of the most interesting classes
of solids, relatively little theoretical work aiming at a microscopic
understanding of electron-correlation effects in these systems is
available so far. This is not surprising: high numerical effort is
already required here for accurately describing ground-state properties
at the Hartree-Fock (HF) independent-particle level.
Recently, results
on NiO \cite{Towler,doll3}
and on the rutile ${\rm TiO_2}$ crystal \cite{silvi,reinh} were published.
Correlation effects in TiO$_2$ have been studied only implicitly, at the
density-functional level \cite{reinh,soran,glass}.

Rutile is one of the experimentally found
modifications of TiO$_2$. The investigation of the relative stability
compared to other phases such as anatase, brookite  or TiO$_2$(B)
is another interesting subject \cite{joenic}.
Rutile is
well studied by experimentalists; the
structure was precisely determined by using X-ray
and neutron diffraction (for a discussion of the experimental
results see
Ref. \onlinecite{glass} and references therein).
The experimental lattice constant is close to
values obtained in
{\em ab-initio}  self-consistent field (SCF)
calculations \cite{reinh}, whereas the difference
between SCF
results and experiment for the compressibility is large.
This contradiction
makes the task of studying electron correlations in the rutile crystal
interesting.

From the methodological point of view
the rutile crystal is  a next logical step in the application of a
correlation treatment within the
so-called "scheme of local increments"  \cite{incre1,incre2}.
The validity of this approach was
successfully tested for covalently bonded solids
like diamond, graphite and many typical semiconductors
\cite{incre2,incre3,beat2,simon,beat1}.
For purely ionic crystals the scheme works also well
(see references for MgO, CaO, NiO and alkali halides
\cite{doll1,doll2,doll3,doll4}).
II-VI semiconductors have been investigated as an example of partly ionic,
partly covalent crystals \cite{martin}. With respect to this, rutile is
another excellent object for testing.

In the present short contribution we report on {\em ab-initio}
correlation
calculations for the cohesive energy, lattice
constant and compressibility of rutile.
The paper is organized as follows.
In the second chapter we outline the computational method and
describe our results. We also discuss individual correlation-energy increments
and address the question of their transferability when approximating
the infinite crystal by embedded clusters. A discussion and a short summary
are presented in the last chapter.

\section{Computational method}

The rutile structure is tetragonal (nonsymmorphic space group
$P4_2/mnm$) with two titanium atoms and four oxygens per primitive
unit cell (see Fig. \ref{tio2bild}). 
Ti atoms are located at positions $(0, 0, 0)$ and
$(1/2, 1/2, 1/2)$ and oxygens at  positions $(\pm x, \pm x, 0)$
and $(1/2 \pm x, 1/2 \mp x, 0)$. The lattice constants
extrapolated to zero temperature \cite{crc}
are $a= 4.592$ {\AA} and $c=2.958$ {\AA},
the  dimensionless coordinate $x= 0.3048$.
The experimental value for $\theta$ is  $98.8^\circ$.
The bonding and charge distribution in the rutile crystal
have been discussed in  Ref. \onlinecite{soran}.
The SCF Mulliken population analysis
\cite{silvi,reinh} shows that the excess negative charge on oxygen is
about $-1.4 e$.
On Ti, in addition to the closed (but easily deformable)
argon-like core the population analysis gives 1.2
valence electrons in the $d$ shell, whereas the
4$s$ shell is not occupied \cite{silvi,reinh}.

After this preliminary discussion, we briefly review our theoretical
approach for including electron correlation on top of a crystal SCF
calculation.
The main idea is to expand the total correlation energy, $E_{corr}$,
of the crystal in terms of local correlation-energy increments.
Details have been
described in previous papers
\cite{incre1,incre2,incre3,beat2,doll1}.
Essentially, we use a Bethe-Goldstone-like hierarchy of the type
\begin{eqnarray}
E_{corr} =  \sum_A \varepsilon(A) + \frac{1}{2}
\sum_{A,B}\Delta \varepsilon(AB) + \\ \nonumber
\frac{1}{3!}\sum_{A,B,C}\Delta \varepsilon(ABC) + \ldots
\end{eqnarray}
where $A, B, C, \ldots$  denote groups of occupied localized
orbitals on atoms $A, B, C, \ldots $, respectively (we use the same  symbol
for an atom and for the group of the localized orbitals on $A$).
The quantity $\varepsilon(A)$ (a one-body increment)
denotes the (local) correlation energy of the crystal
in which only the orbitals of group $A$ are correlated.
The two-body increment $ \Delta \varepsilon(AB)$
is defined as the non-additive part of
correlations arising from simultaneously correlating the groups $A$ and $B$:
\begin{equation}
\Delta \varepsilon(AB) = \varepsilon(AB)- \varepsilon(A)- \varepsilon(B)
\end{equation}
Similarly three-body increments can be introduced as follows:
\begin{eqnarray}
\Delta \varepsilon(ABC) = \varepsilon(ABC)-
\Delta \varepsilon(AB)-  \\ \nonumber
\Delta \varepsilon(AC)-
\Delta \varepsilon(BC)-
\varepsilon(A)- \varepsilon(B) - \varepsilon(C)
\end{eqnarray}
Even if one should want to treat these formulae as
purely phenomenological ones, still
the expansion (if carried out to infinity)
is exact which is clear upon  inspection.
(For a more sophisticated derivation, cf.\ Ref.
 \onlinecite{tom}.)
For  practical applications, however, some approximations are inevitable.
The first approximation consists in truncating the infinite series.
The data for various systems
\cite{incre1,incre2,incre3,beat2,simon,beat1,martin,doll3,doll1,doll2,doll4}
show that the expansion is quickly convergent. The three-body
increments were found to be almost negligible (for the bulk of
practical applications).
Moreover, two-body increments
were found to decay rapidly for larger distances.
The second (more serious) approximation for the incremental
expansion consists of replacing the
infinite crystal with finite embedded clusters for the purpose of
determining individual increments.
This approximation uses the local nature of electron correlation
\cite{fulde} and was found to work well.
Note that we restrict the local cluster treatment to correlation effects
only --- SCF interactions are long-range, and
a calculation
involving the whole (infinite) lattice is mandatory here.
This is possible with the CRYSTAL
program package \cite{crys1,crys2}.

Applying the combined scheme just described to TiO$_2$, 
one
obtains the total energy functional
\begin{equation}
E_{tot}(a,c,x) = E_{scf}(a,c,x) + E_{corr}(a,c,x)
\end{equation}
as a function of the lattice parameters $a,c,x$ or $r_1$, $r_2$,
$\Theta$ (in internal coordinates).
The crystal compressibility $B$, at zero temperature, is defined as
\begin{equation}
B = V \frac{\partial^2 E_0(V)}{\partial^2 V}
\end{equation}
with $V$ denoting the volume of the unit cell and  $E_0(V)$ the
{\em conditional} minimum of $E_{tot}(a,c,x)$ for {\em fixed constant
volume}. Using $V = a^2 c$ and expressing $E_{tot}(a,V/a^2,x)$  as a
function of two independent variables $a, x$,
the  compressibility $B$ can be obtained.
This method is usually not applied to noncubic crystals, but instead
the minimum position at $a_0, c_0, x_0$ is approached
by a conjugated gradient technique. This reduces the number of data
points, and the compressibility can only be obtained then
by applying an empirical ansatz such as the Murnaghan equation of state
(cf.\ e.g.\ Ref. \onlinecite{reinh}).
 Another possibility is to define an
{\em artificial} isotropic compressibility \cite{reinh}  $B_{ISO}$
where one (incorrectly) assumes that under compression
$x$ remains constant and $a$ and $c$ scale isotropically.

\subsection{Basis sets}

\noindent
{\em SCF calculations:} For the titanium atom, a relativistic energy-consistent
12-valence-electron pseudopotential \cite{ecp}, together with
a 411/411/41 basis set
\cite{reinh} was used.
For oxygen we chose the basis set given by
Caus\`a et al \cite{causa}.
To calculate the energy of the free
atoms, diffuse functions  cannot be omitted (let us remind that they
must be omitted in CRYSTAL calculations \cite{crys1,crys2}).
Thus we appropriately supplemented the basis sets for this 
purpose \cite{atombasis}. Our SCF cohesive energy is smaller compared to
that of Ref. \onlinecite{reinh} probably just because of the lack of
diffuse functions for the free Ti atom
in Ref. \onlinecite{reinh}. Of course,
this omission affected only the cohesive energy.

\noindent
{\em Correlation calculations:} For oxygen we used
Dunning's correlation-consistent augmented valence triple
zeta $[5s4p3d2f]$ basis set  \cite{dunn2}.
For Ti,  our starting point was a $[6s5p3d]$ basis set
optimized for the
pseudopotential of Ref.\ \onlinecite{ecp}.
We decontracted one $p$ and one $d$ function, ending up
with a $[6s6p4d]$ set which was supplemented with
 $2f1g$
polarization exponents optimized in CCSD calculations
for the free ground-state atom
($f$ exponents are 2.45, 0.766, and the $g$-exponent is 2.132).

\subsection{SCF calculations}

With the CRYSTAL95 program package \cite{crys1,crys2},
SCF energies for 100 different sets of  randomly chosen $a, c, x$ values were 
calculated.
A region around the minimum was  covered  more
densely.  The computational  results were fitted
by a 3-variable polynomial of third order plus a few quartic terms
which turned out to be significant. (Typically about twenty constants
to fit.)

Some final results of SCF calculations are collected in Table
I. We recover at the SCF level about 57 \% (834 mH) of the experimental 
cohesive energy. The lattice constant is already close to experiment, but the
bulk modulus turns out to be substantially too high. These results indicate
the importance of taking into account electron correlation.


\subsection{Correlation calculations}

For the correlation calculations, we
applied the coupled-cluster approach with single and double excitations (CCSD)
\cite{abin3} and
included perturbatively triples (CCSD(T))\cite{molpro1} as implemented in the
program package MOLPRO
\cite{molpro2,molpro3}.
Within the incremental scheme, we studied
one-, two- and three-atom clusters.
(For the three-atom clusters, we performed CCSD calculations only.)
The clusters were embedded in a large slab of Madelung
point charges ($ 7 \times 7 \times 9$ unit cells, charges +4 and -2, 
respectively),
similarly as described in Refs. \cite{doll3,doll2,doll1,doll4}. 
An exception are the Ti ions 
nearest to the cluster atoms: in this case, Ti$^{4+}$ pseudopotentials
\cite{hay} were used
instead of the bare point charges, in order to
simulate
the Pauli repulsion on the O$^{2-}$ electrons of the inner cluster.
{\em The not fully ionic character of the system turns out to be
the basic difficulty.}
When doing computations on finite clusters, only an
integer electron number is allowed and the atomic populations obtained from
the CRYSTAL calculation cannot easily be reproduced as is the case for
perfectly ionic systems.
We are forced to
assign to the clusters considered  the same total electron
charges as in hypothetical purely ionic Ti$^{4+}$, O$^{2-}$
rutile. Of course, this is no issue in large clusters, but may be critical
for small ones. In order to control the quality of this approximation, we 
compare the
results for the individual increments taken from clusters with one, two,
and three explicitly described ions
(see Table II).

The bulk of our correlation-energy calculations
(whose results appear in Table I) was done for
embedded clusters with one and two explicitly treated
atoms.
The maximum distance between the atoms within these
clusters was up to 8 atomic units.
These clusters (altogether 13 in number)
provide us with the most significant one-body and two-body
contributions to $E_{corr}$.
For 5 different geometries, all these increments were calculated.
For 15 other geometries, the 8 most important increments were explicitly
calculated whereas  the remaining 5 least important
(contributions below 1 mH  )
were obtained by
interpolation.
To estimate the importance of three-body contributions,
we also calculated
the biggest three-body $O-Ti-O$ increments for one geometry, and they
were found to be of the order of
1 mH.

A very important question is
the transferability of these increments from one type
of cluster to the other (see Table II).
In every case, our increments refer to localized orbital groups
which can be formally attributed to O$^{2-}$ and Ti$^{4+}$ ions,
respectively, but due to partially covalent bonding in TiO$_2$,
especially the former orbital group becomes more delocalized when
going to larger clusters. Specifically, we see that 
the oxygen correlation energy increases in 
magnitude when taken from a cluster with three explicitly described ions
(one titanium ion, two oxygen ions, to be consistent with the formula unit) 
instead of out of a cluster where only the oxygen ion has basis functions.
We explain this with a better description of the diffuse tail of the
oxygen charge, as the charge is allowed to flow to the titanium ion. The 
more diffuse ion has lower-lying excitations which explains the change
of the correlation energy. In the same way, we can compare the titanium
correlation energy of a cluster where only one titanium ion has basis 
functions to that of a cluster with three explicitly described ions.
We find that the titanium correlation energy is reduced in magnitude.
This is due to the oxygen charge which has flown to the titanium ion and
has led to a $d$ occupancy. Excitations of the $3s$ and $3p$ electrons
into $d$ orbitals 
are now (partially) forbidden, i.e., we have an exclusion effect. In total, 
both effects
almost cancel for the one-body increments ($|\Delta \varepsilon(O)|$ 
increases by 5 mH, to be multiplied with a weight factor of 4, 
($|\Delta \varepsilon(Ti)|$ decreases by 11 mH, to be multiplied with 2).
For the two-body increments, a partial cancellation with increasing cluster 
size
takes place, too: O-O increments are enhanced by $\sim$4 mH when bridging Ti 
atoms
are taken into account, while Ti-O increments are reduced in magnitude by 
$\sim$3 mH in the TiO$_2$ unit \cite{add}.

As a whole, we find that the transferability is reasonable but
rather poor in comparison to purely ionic crystals
where cluster charges are strictly confined (see Table II).
This is not unexpected as the {\em perfect-ion approximation}
is broken in a different way
for each of the clusters studied.
To overcome this transferability 
problem, the only way would be to further improve the
cluster surroundings by using more than three explicitly described ions
with high quality basis set
(ideally in multiples of the formula unit),
which is presently not possible because of the steeply increasing 
computational effort.


The correlation energies obtained this way were fitted for 20 geometries
by a second-order polynomial (ten parameters), analogously as previously
described
for SCF energies. Adding the two resulting formulae,
we obtained 
$E_{tot}(a,c,x)$ so that we could perform the calculation of
physical constants which include correlation effects (see Table I).

\section{Discussion and summary}

As SCF and SCF+CCSD(T) lattice constants are not very different and both
rather close to the experimental values we conclude
that electron correlations do not influence the crystal geometry
in a major way.
The technical explanation is provided by studying the coefficients
of the analytic (fitted) formula of $E_{corr}(r_1,r_2,\theta)$,
in atomic units, at the CCSD(T) level:
\begin {eqnarray}
E_{corr}(r_1,r_2,\theta) \approx
-2.1241
-0.0121 \Delta r_1 \\ \nonumber
- 0.0067 \Delta r_2 
- 0.0322 \Delta \theta   \\  \nonumber
+ 0.2570 (\Delta r_1)^2
- 1.5730 (\Delta r_2)^2
- 1.2490 (\Delta \theta)^2   \\ \nonumber
- 3.0459 \Delta r_1 \Delta r_2
+ 1.3580 \Delta r_1 \Delta \theta
- 0.0875 \Delta r_2 \Delta \theta
\end{eqnarray}
The normalized dimensionless expansion variables are defined
as $\Delta r_1 =   (r_1 -r_{10})/r_{1exp}$ where   $r_{10}$ is
the SCF equilibrium value of the internal variable $r_{1}$
and $r_{1exp}$ is the corresponding experimental value. Analogous
expressions hold for $\Delta r_2$ and for $\Delta \theta$.
The coefficients of the linear terms are
much smaller than the coefficients of the quadratic terms.
This explains why the compressibility
is strongly renormalized
by correlations as it depends primarily on
the coefficients of the quadratic term.

To understand this situation
it is necessary to study how
 individual increments change when the lattice constants change.
To gain  a qualitative understanding it is enough
to study the biggest contributions: the one-body increment for oxygen and
the two-body nn Ti-O increments.
(Note that the one-body increment for Ti can be considered as constant,
to a good approximation).
In the following we will describe several
competing mechanisms
which cause linear changes of the bulk correlation energy to be
small near the SCF minimum.

Let us start with oxygen (one-body increment for oxygen).
Its change as a function of the lattice constant
is in part of intraatomic origin and in part of electrostatic
origin.
First let us consider the intraatomic part. As found and explained in
earlier work
\cite{doll1,doll2,doll3}, excitations cost less energy when the lattice
constant (and the volume of the quantum well
enclosing the oxygen ion) increases; in that case,
the magnitude of the correlation energy increases, too.
There is an opposite trend connected with the permanent electric
field $\vec E$ at the oxygen site. As the lattice expands, the
Madelung field decreases, and
the influence of correlation on the static polarization of the oxygen ions,
$\frac{1}{2}(\alpha_{SCF}-\alpha_{corr}) \vec E^2$
(with polarizabilities $\alpha_{SCF}$ and $\alpha_{corr}$ at SCF and
correlated levels, respectively), decreases in parallel.
(Note that due to symmetry there is no static electric field at the site of the
Ti ions).
A competition between the two effects just described apparently
leads to a relatively small change in the total oxygen correlation energy as
a function of the lattice constant. 

Thus, the most important source of variability are two-body increments.
For the two-body nn Ti-O increments, we find two competing
effects again.
To fix the attention let us provide  two greatly simplified formulae
(atomic units):
\begin{eqnarray}
\Delta\varepsilon_{api}  \approx  -0.027 + 0.083 \Delta r_1  - 0.039 \Delta
r_2  - 0.036 \Delta \theta
\cdots   \\ \nonumber
\Delta\varepsilon_{equa}  \approx  -0.026 - 0.046 \Delta r_1
+ 0.018 \Delta r_2  +0.040 \Delta \theta
  \cdots
\end{eqnarray}
where $\Delta \varepsilon_{api}$ and 
$\Delta \varepsilon_{equa}$ are two-body nn Ti-O
increments   for apical and equatorial
oxygen ions, respectively. The large quadratic terms in the above
formulae were not shown as they are irrelevant to the following arguments.
Let us remember that the multiplying weight
factors
(per unit cell)  for
$\Delta \varepsilon_{api}$ and
$\Delta \varepsilon_{equa}$  are 4 and 8, respectively.
Due to the opposite signs and due to the fact that the different linear term
amplitudes are roughly in  1:2 proportion we may conclude that
a small lattice distortion
can change the individual $\Delta \varepsilon$ but
not the sum $ 4 \Delta \varepsilon_{api} + 8 \Delta \varepsilon_{equa}$
which is roughly constant (but only in linear approximation).
Thus, $ 4 \Delta \varepsilon_{api} + 8 \Delta \varepsilon_{equa}$
(and consequently the total correlation energy as well)
is characterized by {\em  small linear terms} and by
{\em large quadratic terms}.
The physical explanation is given by  a competition
between the bonds Ti - apical O and Ti - equatorial O.
Suppose we distort the crystal in such a way
that  only the internal coordinate $r_1$ increases.
The apical oxygens are a little further apart and the
interatomic van der Waals-like correlation energy contained within  
$\Delta \varepsilon_{api}$
decreases in magnitude. At the same time,
the polarizing influence of the Ti ions on the charge clouds of the apical 
oxygen anions
decreases; this leads to a charge displacement from these anions towards 
their other Ti 
neighbours,
with respect to which they are in equatorial positions, and,
as a result,
the electron correlation contained in
$|\Delta \varepsilon_{equa}|$ is getting bigger.
Analogous effects (of opposite sign) arise with a change of $r_2$,
with the exception that the two equatorial Ti neighbours of an oxygen
anion are moved simultaneously, which leads to a smaller prefactor of 
$\Delta r_2$, in the expression for 
$\Delta \varepsilon_{equa}$, as compared to that of $\Delta r_1$ for
$\Delta \varepsilon_{api}$.
 The two discussed types of
correlation-energy change
are opposite and in linear approximation almost cancel.

In conclusion we have shown that electron correlations
do not change the lattice geometry in a major way but are important
for other ground-state properties --- the cohesive energy
and bulk compressibility which we obtained are close to the experimental
values.
In view of these results, we conclude that the application of the
incremental scheme to rutile yields new insights (in spite of the fact that
we were forced to go to approximations which
are problematic from the methodological point of view).
In addition we gained valuable experience about how
to apply this scheme to mixed covalent-ionic systems
with fractional ion charges.

\acknowledgements
We would like to thank Dr.\ M.\ Dolg, of the
MPI-PKS (Dresden), for
valuable discussions.


\newpage

\widetext

\begin{table}
\caption{ Ab initio results for rutile}
\label{table1}
\begin{tabular}{lddddd}
 & SCF \tablenotemark[1]  & SCF \tablenotemark[2] &
SCF present work  & SCF + CCSD(T) &
experiment
\\
\hline
$a_0$\tablenotemark[3]  & 4.559 & 4.555   &  4.529   & 4.548 &  4.592
\\
$c_0$\tablenotemark[3]  & 3.027 &3.024   &  3.088   & 2.993 &  2.958
\\
$x_0$  & 0.3048 &0.3061   &  0.3052  & 0.3046 &   0.3048  \\
$B$\tablenotemark[4] &2.81& 2.79  &  3.04  & 2.36
\tablenotemark[6]  & 2.39 \\
    \\
$B_{ISO}$\tablenotemark[4] &3.09 & 3.08  &  3.15 & 2.45 &       -
\\
$\Delta E_B$\tablenotemark[5]  &  1.025   & 1.175 & 0.834 & 1.422
&  1.470\\
\end{tabular}
\tablenotetext[1]{
Ref. \onlinecite{reinh} data (all-electron calculations)
}
\tablenotetext[2]{
Ref. \onlinecite{reinh} data (pseudopotentials on Ti and O)
}
\tablenotetext[3]{
units are  \AA
}
\tablenotetext[4]{
units are $Mbar$
}
\tablenotetext[5]{
 $\Delta E_B$ is cohesive energy
per unit cell; in Hartree units, including zero point
vibrations with a Debye approximation and a Debye temperature of 530 K
\cite{deby2}; the energies of the free atoms are averaged over $J$ 
with the appropriate experimental spin-orbit contributions \cite{moore})}
\tablenotetext[6]{
The estimated error bar for $B$ is $\pm 0.06$      }
\end{table}

\newpage

\begin{table}
\caption{Selected correlation-energy increments in rutile, from CCSD 
calculations (SCF equilibrium geometry from Ref. 4, 
2nd column in Table I;
atomic units). For the free atoms, we obtain correlation energies of 
-0.1714 H for O, and -0.4040 H for Ti, at the CCSD level.}
\label{table2}
\begin{tabular}{llddd}
 &  cluster & increment   & weight factor \\
 &          &  & per unit cell \\
\hline
$\Delta \varepsilon(O)$ & O                     & -0.2613  &4\\
                        & Ti-O\tablenotemark[1] & -0.2692  &\\
                        & Ti-O\tablenotemark[2] & -0.2692  &\\
                        &  O-O\tablenotemark[3] & -0.2607  &\\
                     &  O-Ti-O\tablenotemark[4] & -0.2667  &\\
                     &  O-Ti-O\tablenotemark[5] & -0.2666  &\\
                     & Ti-O-Ti\tablenotemark[6] & -0.2736  &\\
\hline                                                    
$\Delta \varepsilon(Ti)$ & Ti                   & -0.3117 &2\\
                        & Ti-O\tablenotemark[1] & -0.3054 & \\
                        & Ti-O\tablenotemark[2] & -0.3052 & \\
                     &  O-Ti-O\tablenotemark[4] & -0.3009 & \\
                     &  O-Ti-O\tablenotemark[5] & -0.3006 & \\
                     & Ti-O-Ti\tablenotemark[6] & -0.3062 & \\
\hline
$\Delta \varepsilon(Ti-O_{api})$\tablenotemark[1] 
                        & Ti-O\tablenotemark[1] & -0.0223   &4\\
                     &  O-Ti-O\tablenotemark[4] & -0.0195   &\\
                     & Ti-O-Ti\tablenotemark[6] & -0.0200   &\\
\hline
$\Delta \varepsilon(Ti-O_{equa})$\tablenotemark[2] 
                        & Ti-O\tablenotemark[2] & -0.0239   &8\\
                     &  O-Ti-O\tablenotemark[4] & -0.0206   &\\
                     &  O-Ti-O\tablenotemark[5] & -0.0204   &\\
                     & Ti-O-Ti\tablenotemark[6] & -0.0208   &\\
\hline
$\Delta \varepsilon(Ti-Ti)$\tablenotemark[7] 
                     & Ti-Ti\tablenotemark[2] & -0.0001  &2\\
                    & Ti-O-Ti\tablenotemark[6] & -0.0001  &\\
\hline
$\Delta \varepsilon(O-O)$\tablenotemark[3] 
                        & O-O\tablenotemark[3] & -0.0051   &2\\
                     & O-Ti-O\tablenotemark[5] & -0.0093   &\\
\hline                                                     
$\Delta \varepsilon(O-O)$\tablenotemark[8] 
                        & O-O\tablenotemark[8] & -0.0029   &16\\
                     & O-Ti-O\tablenotemark[4] & -0.0066  &\\
\hline                                                     
$\Delta \varepsilon(O-Ti-O)$\tablenotemark[4] 
                    & O-Ti-O\tablenotemark[4] & +0.0012 &   16\\
\hline
$\Delta \varepsilon(O-Ti-O)$\tablenotemark[5] 
                    & O-Ti-O\tablenotemark[5] & +0.0013 & 4\\
\hline
$\Delta \varepsilon(Ti-O-Ti)$\tablenotemark[6] 
                    & Ti-O-Ti\tablenotemark[6] & +0.0001 &  4\\
\hline
\end{tabular}
\tablenotetext[1]{
Ti -- nearest neighbour apical oxygen (distance - 3.72 a.u.)
}
\tablenotetext[2]{
Ti -- nearest neighbour equatorial oxygen (distance - 3.70 a.u.)
}
\tablenotetext[3]{
nearest neighbour oxygen pair (distance - 4.72 a.u.)
}
\tablenotetext[4]{
equatorial oxygen -- Ti -- apical oxygen (distances: 3.72 and 3.70
a.u.)}
\tablenotetext[5]{
equatorial oxygen -- Ti -- equatorial oxygen (distances: 3.70 and 3.70
a.u.; the O-Ti-O angle is $79^\circ$.)}
\tablenotetext[6]{
Ti - O - Ti  (distances: 3.72 and 3.70
a.u.)
}
\tablenotetext[7]{
nearest neighbour Ti-Ti pair (distance - 5.71 a.u.)
}
\tablenotetext[8]{
next nearest neighbour oxygen pair (distance - 5.25 a.u.)
}
\end{table}

\begin{figure}
\caption{The rutile structure with oxygen (open circles)
and titanium atoms (filled circles).}
\label{tio2bild}
\end{figure}

\begin{references}
\bibitem{atombasis} We added two $sp$ exponents (0.21, 0.07) and one $d$
exponent (0.1) to the titanium basis set, the oxygen basis set was augmented
with two additional $sp$ exponents (0.1, 0.04). The energies of the neutral
atoms in their ground state are -57.72907 H (Ti) and -74.79135 H (O).
\bibitem{add} 
Additional calculations (with somewhat reduced basis sets) for a TiO$_6$ 
cluster
featuring the full crystal coordination of a Ti$^{4+}$-like entity lead to an
even larger reduction of the TiO increment, by $\sim$5 mH, thus still improving
the cancellation with the changes in the O-O increments.
\end{references}
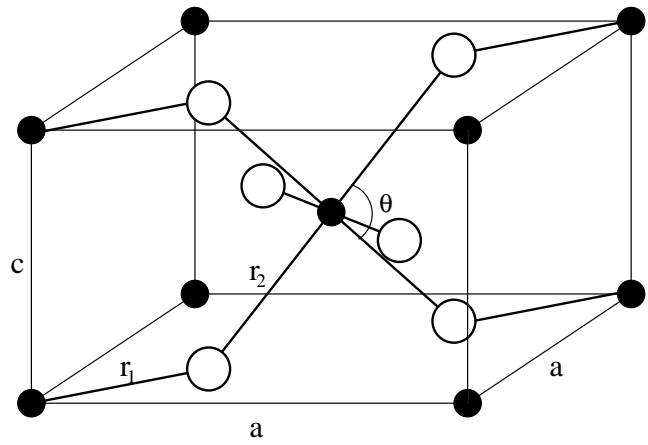
\end{document}